\documentclass{kluwer}
\usepackage{epsfig}
\begin{document}
\begin{article}

\begin{opening}
\title{Numerical Simulations of Interacting Gas-Rich Barred Galaxies}
\author{{\bf I. Berentzen}, C.H. Heller \& K.J. Fricke
         \email{iberent@uni-sw.gwdg.de}}
\institute{Universit{\"a}ts--Sternwarte G{\"o}ttingen, Germany}
\author{E. Athanassoula}
\institute{Observatoire de Marseille, France}

\begin{abstract}

 Using an N--body+SPH code we have performed numerical
 simulations to investigate the dynamical effects 
 of an interaction between an {\em initially} barred galaxy and a small
 spherical companion. 
 In the models described here the small companion passes through the disk of
 the larger galaxy perpendicular to its plane. The impact positions and times
 are varied with respect to the evolutionary phase of the bar and disc.
 The interactions produce expanding ring structures, offset bars, spokes, and
 other asymmetries in the stars and gas. They also affect the
 strength and pattern speed of the bar. 

\end{abstract}
\end{opening}

\section{Introduction}

 The evolution of disc galaxies is driven by both internal and external
 processes. Internal instabilities in the disc often give rise to the formation
 of a bar. Bars are a common feature in disc galaxies and by no means
 an exception, since about $2/3$ of all disc galaxies are believed to harbor a
 bar or oval distortion. Furthermore
 in the last few years near-infrared observations have given clear
 indications of an even higher fraction of barred galaxies.
 The presence of a bar changes both the kinematics and mass (stellar and gas)
 distributions within the disc and can give rise to dynamical resonances
 (e.g., \opencite{Sel93}).
 Similarly, it has become increasingly recognized that interactions 
 with small companions are a frequent and important external agent 
 driving dynamical evolution. Since both these processes are thought
 to be frequent, it is a reasonable assumption that the interaction between a
 barred disc galaxy and a small companion is a common event.

 The study of \inlinecite{Ath97}, using pure N-body
 simulations,  has demonstrated that a sufficiently massive
 companion hitting the inner parts of a barred disc galaxy can
 displace the 
 bar and produce expanding stellar rings. In this work we have included
 the dissipative effects of a gas component.

\section{Numerical Methods}

 For the simulations we use an N-body/SPH code to evolve the stellar and
 gaseous components of the galaxies. A detailed description of the 
 algorithm can be found in \inlinecite{Hel95} and \inlinecite{HeS94}. The
 gravitational forces and neighbor interaction lists were
 computed using the special purpose hardware GRAPE-3AF. The advantage
 of the GRAPE 
 hardware, besides its speed, is that it does not set any
 constraints on the symmetry of the simulations. The adopted units for mass,
 distance, and time are $M\!=\!6\cdot10^{10}$\,M$_{\odot}$, $R\!=\!3$\,kpc, and
 $\tau\!=\!10^7$\,yr, respectively, and the gravitational constant is taken
 as unity.  
 A fixed gravitational softening length of
 $\epsilon\!=\!0.1875$\,kpc is used for all particles.

\subsection{Initial Conditions}
 
 Both the stellar and gaseous discs are initially setup with a truncated
 Kuzmin-Toomre projected radial density profile and a $sech^2$
 vertical distribution. The halo and companion have a
 truncated Plummer profile. The parameters are given in Table\,1
 and were chosen so as to give a bar unstable model. 

 Based on the potential of the disc and relaxed halo we assign velocities
 to the disc particles. The velocity dispersion is derived from the
 local stability 
 criterion and is corrected for asymmetric drift. The resulting rotation curve
 has a maximum of $v_{\rm max}\!\approx\!200$\,km/s and remains flat in
 the outer disc. 

 The initial conditions for the interactions are specified by the impact
 position
 ($x_{\rm i},y_{\rm i}$) with respect to the bar and the impact time
 $t_{\rm i}$. 
 The impact velocity $v_{\rm i}$ of the companion is chosen to be roughly equal
 to $4\,v_{\rm esc}$, with
 $v_{\rm esc}$ being the escape velocity from the center of the target
 system. The orbit is 
 integrated backward for a time period of $\Delta t\!=\!30$, which places the
 companion well outside the halo of the disc system. 
 A more complete description of the procedure for setting up the interaction
 configuration can be found in \citeauthor{Ber99a} (1999\,a,b). 
 For the simulations in this paper we have chosen two different impact times,
 one at $t_{\rm i}\!=\!60$ and the other at $t_{\rm i}\!=\!150$. As will be
 shown in
 the next sections, the former corresponds roughly to the time of the
 maximum bar strength and the latter to a time when the bar is
 considerably weaker. We will thus hereafter refer to cases with the first
 impact time as interactions with a strong bar and to cases with the
 second one as interactions with a weak bar.

\begin{table}
\caption{Initial model parameters}
\begin{tabular}{cccrlccc} \hline
         &        &   Type & N$_D$   & M$_D$ & a$_D$ & R$_{\rm cut}$ & z$_0$ \\
\hline
{\bf Disc} & stars  &  KT  & 13\,500 & 0.56  &  1.0  &  5.0        &  0.20 \\
           &   gas  &  KT  & 10\,000 & 0.14  &  1.0  &  5.0        &  0.05 \\
{\bf Halo} & stars  & PL & 32\,500 & 1.30 & 5.0 & 10.0          & $\cdots$\\
{\bf Companion}& stars & PL & 10\,000 & 0.40 & 0.195 & 3.0     & $\cdots$ \\
\hline
\end{tabular}
\end{table}

\section{Morphological Evolution}

\subsection{Isolated Galaxy}

 The model was constructed so as to be globally unstable to non-axisym\-metric
 perturbations and forms a large-scale bar in a few disc rotations
 ($t_{\rm rot}$, measured at disc half-mass radius).
 The bar reaches
 its maximum strength at $t\!=\!60$ or some $20\,t_{\rm rot}$. At this time the
 bar has a major axis length of $a\!=\!6$\,kpc and an axial ratio of
 $\sim\!3\!:\!1$. Both stellar and gaseous trailing spiral arms emerge from the end of the bar. 
 While the stellar arms slowly dissolve and are hardly
 visible at $t\!=\!100$, large spiral features in the gas persist throughout
 the run. The gas also forms straight off-set
 shocks at the leading edges of the stellar bar .

 Due to the gravitational torque of the bar, the gas is driven towards the center
 and within two bar rotations some $50$ percent of the total gas mass collects
 in an oval nuclear disc elongated along the bar. As a result of the
 growing mass concentration at the center the amplitude of the bar decreases, 
 but then settles down to a quasi-steady state as the initial burst of inflow
 slows down.

 \begin{figure}
 \centerline{\epsfig{file=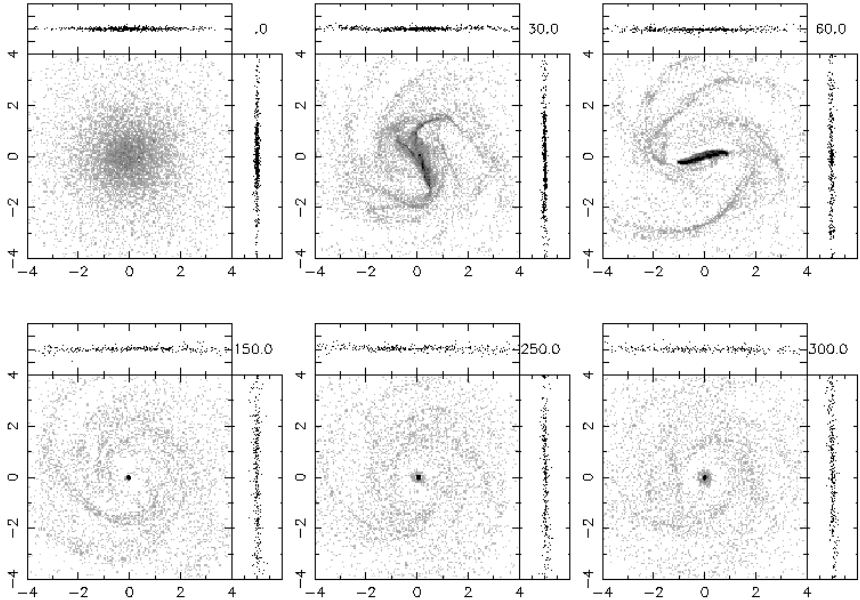, width=12cm}}
 \caption{Isolated model. Shown is the evolution of the gaseous disc only.}
 \end{figure}

 The inner regions are connected with the outer disc through two trailing spiral
 arms, which by way of their shocks feed material inward to the growing central
 disc. Around the nuclear disc forms
 another disc which is oriented perpendicular to the bar major axis. This
 indicates the presence of an ILR, i.e. the existence of an $x_2$
 orbit family. Outside the  
 central region there is a deficiency of gas out to a radius of 3\,kpc,
 where a gas ring forms close to the UHR.  A detailed discussion on orbits
 and resonances in isolated gas-rich barred galaxies can be found in 
 \inlinecite{Ber98}.
 
\subsection{Central Passage}

 In the model with a strong bar both the stellar and the gaseous
 components of the bar get torn apart into two
 fragments. The stellar fragments merge again in about
 $3\,t_{\rm rot}$. The gaseous  
 fragments do not merge as fast, but flow back and forth inside the bar
 potential avoiding the center for some $6\,t_{\rm rot}$ and appear as two
 separate nuclei, 
 until they finally merge to form a dense nuclear disc.  A lower density, more
 extended circumnuclear disc forms around the denser nuclear disc as
 in the isolated model, but with a much larger radius.

 \begin{figure}
 \centerline{\epsfig{file=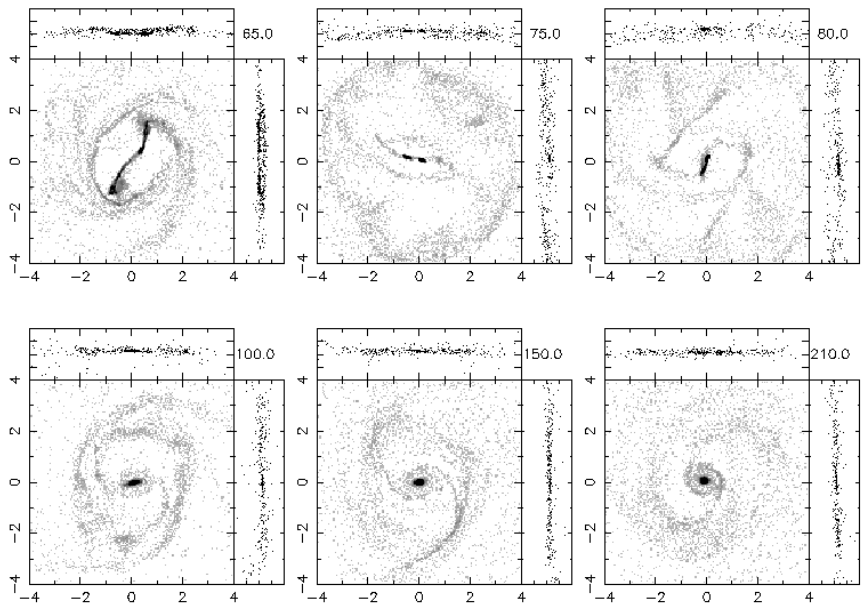, width=12cm}}
 \caption{Central impact with strong bar. Shown is the evolution of the
 gaseous disc only.}
 \end{figure}

 The central impact also creates ring-like expanding density waves in
 the disc, similarly to what happens for non-barred galaxies (\opencite{LT76}).
 The first ring reaches as far as half the
 corotation radius and then fragments, with much of its mass flowing
 back toward the inner few kpc. 
 A second ring propagating outwards follows the first, with trailing
 spokes forming between the two. For the weak bar the rings are almost perfectly
 circular, while for the strong bar they are more asymmetric. One could
 expect such density enhancements to be accompanied by the formation of
 bright young blue stars and H{\rm II} regions.

 The expansion velocity of the first ring drops gradually from 88\,km/s
 to roughly 35\,km/s. The second ring starts with a lower expansion
 velocity of 29\,km/s, 
 decreasing further to 12\,km/s, which is the sound speed of the isothermal
 gas in the model.
 With the passage of the two rings the stellar disc is heated up, preventing
 the formation of a third stellar ring.
 In contrast, more than two rings form in the gas, but from the third
 onwards they are very weak and
 dissolve quickly, since they are not supported by an accompanying 
 stellar density enhancement.  These rings eventually merge with the 
 existing spiral arm segments. 

\subsection{Minor Axis Passage}

 In these simulations the companion hits the disc along the bar minor axis at
 a distance of approximately $r_{\rm i}\!=\!3.0$\,kpc. In both strong and
 weak bar models the stellar bar gets displaced from the center immediately
 following the impact to a distance of approximately 3.0\,kpc in the direction
 of the impact point.
 This offset lasts for a period of roughly $\Delta t\!=\! 0.6$\,Gyr.

 An expanding circular ring is formed at the impact point, which 
 upon encountering the bar and spiral arms becomes distorted.
 The ring expands more freely on the side of the bar which is opposite
 the impact point and also 
 maintains longer its circular shape, before merging with a spiral arm.
 This merged ring/arm then opens up forming a large spiral feature.
 On its other side, the ring stays close to the bar. Also a new spiral
 arm segment forms, connected to one end of the still displaced bar.   

 At the time when the bar is again nearly centered a second ring
 starts to form, expanding outwards from the impact point.  This
 ring merges with the newly formed spiral arm segment.  Between this feature
 and the large spiral arm there is a noticeable deficiency of both
 stars and gas.

 Similar to the central impact, additional rings form at later times
 and merge with the many spiral arms that are present, forming spiral 
 segments which wind up and dissolve over time.  
 The final gas morphology is similar to that of the isolated case, 
 even though the stellar bar has been destroyed.

 \begin{figure}
 \centerline{\epsfig{file=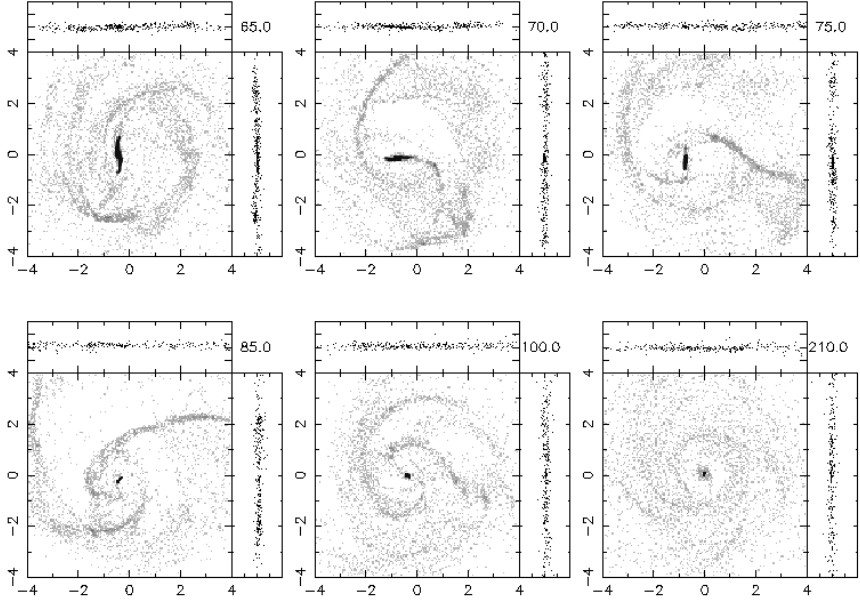, width=12cm}}
 \caption{Major axis impact with strong bar. Shown is the evolution of the
 gaseous disc only.}
 \end{figure}

\subsection{Major Axis Passage}
 
 In these simulations the companion hits the disc on the bar major axis 
 near corotation. Immediately following the passage of the companion
 through the disc plane, the stellar bar is displaced from the center to a
 distance of 2.4\,kpc in the direction of the impact point .

 In general we find the same kind of evolution as in the minor axis
 impact, though the lengths and orientations of the generated features
 differ.  The most significant difference is that the bar survives
 the major axis impact.  In fact the evolution of the bar strength 
 follows closely that of the isolated case.  Again, the final
 gas configuration is similar to the other models.

\section{Evolution of the bar}
\label{bar_evol}

\subsection{Amplitude of the bar}

 We have measured the bar strength as defined by the amplitude of the $m\!=\!2$
 Fourier component of the stellar mass distribution in the disc. For models in
 which the bar gets displaced we trace the center of mass of the stellar bar
 and measure its amplitude within a radius of 3\,kpc from this
 center. The results 
 are shown in figure~2\,(a) and (b).

 \begin{figure}[b]
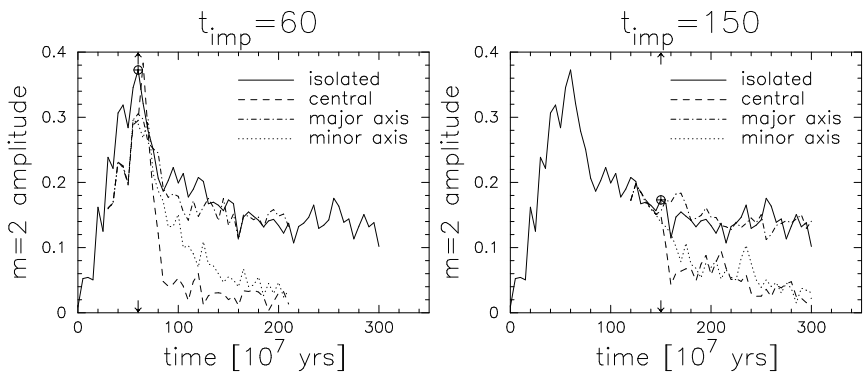

 \centerline{\epsfig{file=figure4a.eps, width=4.8cm, angle=-90} 
             \epsfig{file=figure4b.eps, width=4.8cm, angle=-90}
 }
 \caption{Evolution of bar strength.  Models at impact times of
 (a) $t=60$ and (b) $t=150$ are shown.}
 \end{figure}

 With the central passage the bar is destroyed immediately following the
 impact of the companion, independent of the bar strength.
 In the model with the minor axis passage the bar is also destroyed, but its
 amplitude decreases more slowly over time 
 than in the central passage. Impacts along the bar
 major axis
 show the same evolution of the bar strength as in the isolated model.

\subsection{Pattern speed of the bar}

 In the isolated model the pattern speed increases linearly with time
 during the period in which gas is flowing inward. Once the inflow stops
 the bar quickly settles down to a quasi-steady state and
 rotates with the constant rate of $\Omega_p\!=\!0.3\tau^{-1}$.

 In the case of the central passage, the bar is quickly weakened and it
 is difficult to follow the pattern speed for more than a few disc
 rotations after the impact.  However, the general trend is that of
 a slowing down of the bar in both the strong and weak bar models. 
 
 In the strong bar minor axis passage the pattern speed
 increases with time, converging to a value of $\Omega_p\!=\!0.5$.
 In contrast, the weak bar minor axis passage shows little variation of the
 pattern speed.  The major axis passages show the same evolution
 of pattern speed as in the isolated model.

\section{Summary}

 Using N-body/SPH simulations we have examined the effect of a small
 spherical companion passing through the disc of a gas-rich 
 barred galaxy. The
 impact position has been varied with respect to the bar and two different
 impact times have been chosen, corresponding to weak and strong bar cases.
 
 The interaction produces characteristic features like expanding rings, 
 spokes, and an off-centered bar, similar to what has been found in pure
 N-body simulations. The offset passages lead to a displacement of the bar
 from the center which lasts for approximately 0.6\,Gyr. 

 While the morphology for several disc rotation times following
 the impact is drastically altered by the interaction, the overall final
 state of the bar and disc is similar to that of the isolated case.
 However, the interaction does change the detailed structure of the gas
 distribution in the central region and signs of an interaction in the form
 of asymmetries are left in the outer disc.
 The bar is not necessarily destroyed by the interaction, even though being
 temporarily displaced from the center.

\begin{acknowledgements}
 We would like to thank Albert Bosma for interesting discussions, and
 J.C. Lambert and C. Theis for their computer assistance. 
 The GRAPE 3Af special purpose computer in Kiel was financed under DFG grant
 Sp 345/5-1, 5-2.
 I.B. acknowledges support from DFG grant Fr 325/48-1, 48-2.
 C.H. acknowledges support from DFG grant Fr 325/39-1, 39-2. 
 E.A would also like to thank the IGRAP, the
 INSU/CNRS and the University of Aix-Marseille I for funds to develop
 the grape facilities used for part of the calculations in this paper.
\end{acknowledgements}

\end{article}


\begin{thebibliography}{}
\bibitem[\protect\citeauthoryear{Athanassoula et al.}{1997}]{Ath97}
 Athanassoula E., Puerari I., Bosma A.: 1997,
 {\it MNRAS\/ }{\bf 286}, 284
\bibitem[\protect\citeauthoryear{Berentzen et al.}{1998}]{Ber98}
 Berentzen, I., Heller, C.H., Shlosman, I., Fricke, K.J.: 1998,
 {\it MNRAS\/ }{\bf 300}, 49
\bibitem[\protect\citeauthoryear{Berentzen et al.}{1999a}]{Ber99a}
 Berentzen, I., Heller, C.H., Athanassoula, E., Fricke, K.J.: 1999,
 in: {\em Dynamics of Galaxies and Galactic Nuclei}, SFB 328 Workshop,
 eds. W.J.  Duschl \& C. Einsel (Vol.2 of ITA series), 1999\,a, {\em in press.}
\bibitem[\protect\citeauthoryear{Berentzen et al.}{1999}]{Ber99b}
 Berentzen, I., Heller, C.H., Athanassoula, E., Fricke, K.J.: 1999\,b,
 {\it in prep.}
\bibitem[\protect\citeauthoryear{Heller}{1995}]{Hel95}
 Heller, C.H.: 1995,
 {\it ApJ\/ }{\bf 455}, 252
\bibitem[\protect\citeauthoryear{Heller \& Shlosman}{1994}]{HeS94}
 Heller, C.H., Shlosman, I.: 1994
 {\it ApJ\/ }{\bf 424}, 84
\bibitem[\protect\citeauthoryear{Lynds \& Toomre}{1976}]{LT76}
 Lynds, R., Toomre, A.: 1976
 {\it ApJ\/ }{\bf 209}, 382
\bibitem[\protect\citeauthoryear{Sellwood \& Wilkinson}{1993}]{Sel93}
 Sellwood, J.A., Wilkinson, A.: 1993,
 {\it Rep. Prog. Phys.\/ }{\bf 56}, 173
\end{thebibliography}
\end{document}